\documentclass[sigconf,screen]{acmart}

\usepackage{amsmath,amsfonts}
\usepackage{algorithmic}
\usepackage{graphicx}
\usepackage{textcomp}
\usepackage{xcolor}
\usepackage{mathtools}

\usepackage{url}

\usepackage[binary-units,per-mode=symbol,retain-explicit-plus]{siunitx}
\def\defaultDigitGrouping{5}
\usepackage{booktabs}
\usepackage[acronym]{glossaries} 
\glsdisablehyper 
\usepackage{mdwlist}	

\usepackage{hyperref}

\newacronym{alu}{ALU}{arithmetic logic unit}
\newacronym{asic}{ASIC}{application-specific integrated circuit}
\newacronym{asip}{ASIP}{application-specific instruction-set processor}
\newacronym{cgra}{CGRA}{coarse-grained reconfigurable array}
\newacronym{cpu}{CPU}{central processing unit}
\newacronym{cvd}{CVD}{cardiovascular disease}
\newacronym{dma}{DMA}{direct memory access}
\newacronym{dsip}{DSIP}{domain-specific instruction-set processor}
\newacronym{ecg}{ECG}{electrocardiogram}
\newacronym{fft}{FFT}{fast Fourier transform}
\newacronym{fir}{FIR}{finite impulse response}
\newacronym{fu}{FU}{functional unit}
\newacronym{ipc}{IPC}{instructions-per-cycle}
\newacronym{isa}{ISA}{instruction set architecture}
\newacronym{lsu}{LSU}{load-store unit}
\newacronym{mf}{MF}{morphological filtering}
\newacronym{ncd}{NCD}{noncommunicable disease}
\newacronym{pc}{PC}{program counter}
\newacronym{pe}{PE}{processing element}
\newacronym{ppg}{PPG}{photoplethysmogram}
\newacronym{rc}{RC}{reconfigurable cell}
\newacronym{rsp}{RSP}{respiratory}
\newacronym{srf}{SRF}{scalar register file}
\newacronym{relen}{Rel-En}{Relative Energy}
\newacronym{rms}{RMS}{root-mean-square}
\newacronym[longplural={scratchpad memories}]{spm}{SPM}{scratchpad memory}
\newacronym[longplural={static random access memories}]{sram}{SRAM}{static random access memory}
\newacronym{ulp}{ULP}{ultra-low power}
\newacronym{vliw}{VLIW}{very-long instruction word}
\newacronym{vwr}{VWR}{very-wide register}
\newacronym{wsn}{WSN}{wearable sensor node}
\newacronym{mxcu}{MXCU}{multiplexer controller unit}
\newacronym{lcu}{LCU}{loop controller unit}
\newacronym{sc}{SC}{standard cell}
\newacronym{vwr2a}{VWR2A}{Very-Wide-Register Reconfigurable-Array}

\acmYear{2022}
\acmConference[DAC '22]{Proceedings of the 59th
ACM/IEEE Design Automation Conference (DAC)}{July 10--14, 2022}{San
Francisco, CA, USA}

\begin{document}

\title{VWR2A: A Very-Wide-Register Reconfigurable-Array Architecture for Low-Power Embedded Devices}

\thanks{
This work was supported in part by the Swiss NSF ML-Edge Project under Grant Agreement (GA) no. 200020\_182009, in part by the ReSoRT Project funded by Botnar Foundation under GA no. REG-19-019, in part by the ERC Consolidator Grant COMPUSAPIEN under GA no. 725657, and in part by a joint research grant for ESL-EPFL by IMEC}

\author{Beno\^{i}t~W.~Denkinger}
\affiliation{
   \institution{EPFL}
   \country{Switzerland}
}
\author{Miguel~Pe\'{o}n-Quir\'{o}s}
\affiliation{
   \institution{EPFL}
   \country{Switzerland}
}
\author{Mario~Konijnenburg}
\affiliation{
   \institution{IMEC}
   \country{Netherlands}
}
\author{David~Atienza}
\affiliation{
   \institution{EPFL}
   \country{Switzerland}
}
\author{Francky~Catthoor}
\affiliation{
   \institution{IMEC, and KU Leuven}
   \country{Belgium}
}

\begin{abstract}
Edge-computing requires high-performance energy-efficient embedded systems. Fixed-function or custom accelerators, such as FFT or FIR filter engines, are very efficient at implementing a particular functionality for a given set of constraints. However, they are inflexible when facing application-wide optimizations or functionality upgrades. Conversely, programmable cores offer higher flexibility, but often with a penalty in area, performance, and, above all, energy consumption. In this paper, we propose VWR2A, an architecture that integrates high computational density and low power memory structures (i.e., very-wide registers and scratchpad memories). VWR2A narrows the energy gap with similar or better performance on FFT kernels with respect to an FFT accelerator. Moreover, VWR2A flexibility allows to accelerate multiple kernels, resulting in significant energy savings at the application level.
\end{abstract}


\maketitle

\section{Introduction}\label{sec:intro}

Moving the computational load towards the edge reduces communication energy but requires high-performance energy-efficient embedded devices.
Hardware architecture exploration for such devices is an active area of research, as we still need to obtain higher energy efficiencies to enable long-lasting operation between battery recharges.
The inclusion of hardware accelerators in the computing platform has become a standard in recent years.
These accelerators can execute repetitive operations that account for a significant amount of the processing time (i.e., computational kernels) in a more efficient way than a general purpose CPU or generic GPU.

The development of accelerators follows two main trends: custom accelerators, or ASIC, and flexible (programmable) cores, usually considered as DSIP~\cite{fasthuber2013}.
Fixed-function accelerators are often the most efficient way of implementing a particular functionality for a given set of constraints.
They are, in general, not programmable and are thus focused on a single task or a small family of related tasks.
One example of a custom accelerator is the FFT accelerator included in the MUSEIC platform~\cite{museicv3}, which can execute FFTs of different sizes much more efficiently than the platform ARM Cortex-M4 core.
However, it is possible to build programmable cores tailored specifically for algorithms from a given application domain (i.e., DSIP) without becoming general purpose processors.
Traditional knowledge says that these flexible cores are less efficient than fixed-function ones and that the latter should be preferred to improve the energy efficiency of the systems.

In this paper, we show how to move in the flexibility-performance trade-off and build flexible domain-specific cores that close the efficiency gap in terms of energy and performance with respect to custom ones for individual computation kernels.
Moreover, as these cores still have an instruction-set which covers a broad domain, they hence can be used in more parts of the application.
As  consequence, they achieve larger improvements when the complete application --- rather than individual kernels --- is taken into account.

We started from the basis of a \gls{cgra}~\cite{Duch2017}, which offers a high computation density versus control logic, and improved it with a memory architecture based on \glspl{vwr} and wide \glspl{spm}.
To demonstrate the benefits of our architecture, we considered a biosignal-processing embedded system meant for wearable devices~\cite{museicv3}.
Then, we evaluated two typical kernels used in biosignal applications, FFT and FIR filters, and a biosignal application.

The rest of this paper is organized as follows.
First, we discuss related architectures for low-power computing in Section~\ref{sec:related}.
We then present our proposed architecture for ultra-low power programmable accelerators in Section~\ref{sec:cgraArchitecture}.
In Section~\ref{sec:expSetup}, we describe the experimental setup used to evaluate our proposals, whereas in Section~\ref{sec:expResults} we analyze the results obtained.
Finally, in Section~\ref{sec:conclusion} we draw the main conclusions of our study.

\section{Related work}\label{sec:related}

CGRAs have often been used for DSIP architectures as they provide good flexibility with reasonable energy overhead compared to ASICs.
One example is MorphoSys~\cite{morphosys2000}, a system-on-chip composed of a TinyRISC processor and a CGRA.
The processor controls the execution flow of the code while the CGRA executes the most computationally intensive kernels.
The concept of VWR2A is similar to MorphoSys in that it runs besides a low-performance processor.
The goal is to reduce the overall system energy to the minimum.

However, according to the authors of~\cite{ADRES2003}, such architecture is not optimal because the control-intensive code left to the processor significantly reduces the speed-up benefit of the CGRA.
To address this issue, they proposed the ADRES framework~\cite{ADRES2003}: an architectural template composed of a VLIW processor and a CGRA.
The VLIW allows efficient execution of control-intensive code, and the CGRA still accelerates the computation-intensive kernels.
It is focused on increasing the performance, but not on the low energy consumption.
Whereas in VWR2A, we propose to improve the typical CGRA architecture to execute the intensive-control code.
This allows complete applications to be executed on VWR2A, removing the need for an advanced general-purpose processor like the VLIW of the ADRES platform, thus reducing the overall system energy consumption.

VWR2A integrates low-power memory structures.
The first one is a dedicated \gls{spm} tuned for our biosignal kernel target.
VWR2A is meant to be integrated inside a platform and needs to have access to the system memory, usually through the system bus.
Therefore, the performance of algorithms with many data accesses is dependent on the system bus latency and bandwidth, which can negatively impact the overall performance.
Moreover, data access through the system bus is costly in energy and can be reduced with a judicious memory hierarchy design.
A typical approach is to use caches, like in MorphoSys, but they incur a significant energy penalty due to their inherent control overhead.
\glspl{spm} offer similar performance at lower energy as the control is moved to the software side~\cite{Banakar2002}.

Second, we propose the use of \glspl{vwr} in replacement of the traditional register file for data.
\glspl{vwr} have been introduced in the FEENECS template of~\cite{raghavan2007} and~\cite{b:catthoor2010} as a better alternative, in terms of energy, to standard multi-ported register files.
The first reason is that the cells of the \glspl{vwr} are single-ported, while those of the register files are multi-ported.
Second, their wide interface still allows multiple words to be loaded at once, which leads to a lower overall energy per word access than traditional register files.
To fully benefit from the \glspl{vwr}, the background memory --- the \gls{spm} in our case --- needs to match the \gls{vwr} width, allowing to fill the \gls{vwr} in one cycle.
This effectively makes the \gls{vwr} interface asymmetric with respect to the \gls{spm} side and the datapath side.
The authors of~\cite{raghavan2007} and~\cite{b:catthoor2010} also show that such design is better for place and route, as both memories (i.e., the \gls{spm} and the \gls{vwr}) can be aligned, and the wire length of the most active connection is reduced to a minimum.
Indeed, only the outputs of the multiplexers switch in every cycle, not the outputs of the \glspl{vwr} themselves, reducing the energy consumption.

Several previous works on CGRAs focus on the interconnection scheme of the \glspl{rc}, for example, to provide communication with distant neighbors.
This paper does not consider advanced interconnections between the \glspl{rc} because they represent a significant energy overhead.
Limiting the interconnection to the neighbor cells reduces the wire length, which is essential for an ultra-low power architecture using scaled technologies, without significantly impacting performance.

In this paper, we focus on the novel architectural features of VWR2A.
We integrated it into a specific platform to demonstrate its performance, but it could be integrated into any platform.
CGRA compilers are also an active area of research, but we did not adapt any existing solution to our specific architecture.
We have currently mapped the code manually on VWR2A.

\section{Architecture for ultra-low power programmable cores}
\label{sec:cgraArchitecture}

\begin{figure*}[tp]
	\centering
	\includegraphics[width=0.96\textwidth]{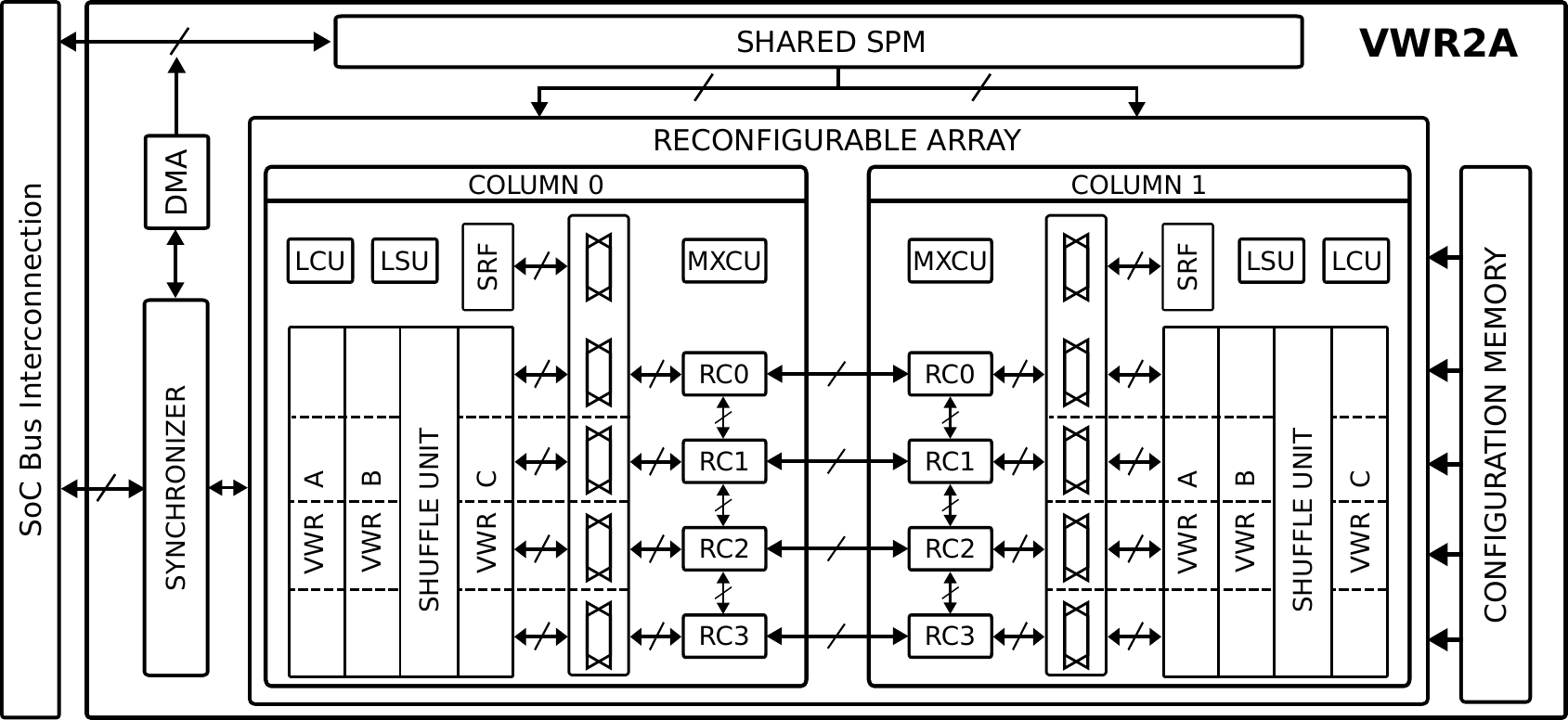}
	\caption{VWR2A architecture diagram}
	\label{fig:vwr2a_architecture}
\end{figure*}

Fig.~\ref{fig:vwr2a_architecture} shows the VWR2A architecture diagram.
We used the CGRA architecture as a basis as it offers a high computation density versus control logic.
The new features and design choices proposed in this paper have been driven by two criteria: extended application code coverage and low power design.

\subsection{Reconfigurable array}

VWR2A contains a 4x2 array of reconfigurable cells.
To reduce control overhead, the \glspl{rc} are grouped in two columns, where all the \glspl{rc} of a column are synchronized through a shared \gls{pc}.
The columns are independent, allowing two kernels to run in parallel.
Each RC has a program memory of 64 words, which is enough to execute most kernels.
The configuration words are stored in the configuration memory and loaded to the \glspl{rc}' local program memory when a kernel execution starts.
The \gls{cgra} architecture offers a high computation density because the bits of the configuration words (i.e., ``instructions'') correspond directly to the control signals in the cell datapaths, without an actual decoding process.
There is evident parallelism between this architecture, where the \glspl{rc} of a column share a program counter, and a VLIW in which all the execution slots are equivalent.
Indeed, the instructions of the different \glspl{rc} can be fused and considered as a wide (predecoded) instruction word.

Each \gls{rc} of the reconfigurable array contains a small register file (two 32-bit entries) and a 32-bit ALU that can execute typical operations: signed addition, subtraction and multiplication, logical bitwise operations, and logical/arithmetic bit shift.
All operations happen in one clock cycle.
The multiplier has two working modes: a standard mode, where the lowest 32 bits are kept, and a fixed-point mode, where the lower 16 bits are discarded, and the next 32 bits are kept.
This enables single-cycle fixed-point multiplication in 16.15 format and good performance for algorithms that require decimal representation, such as the FFT.
The ALU operands have multiple sources: the \glspl{vwr}, the SRF (see Sec.~\ref{subsec:memOrg}), the \gls{rc} local register file, and the previous-cycle results of neighboring \glspl{rc}.
All the operators in the ALUs implement operand isolation~\cite{p:correale1995} to minimize energy consumption.

\subsection{Ultra-low energy memory organisation}\label{subsec:memOrg}

VWR2A contains a dedicated \SI{32}{\kibi\byte} \gls{spm} shared by all the columns.
The \gls{spm} has a double interface:
on the system side, it has the system bus width.
On the accelerator side, it has the same width as the \glspl{vwr}.
A DMA performs the data transfers between the \gls{spm} and the system memory, while the LSU moves the data between the \gls{spm} and the \glspl{vwr}.
The \glspl{vwr} act as a buffer between the \gls{spm} and the \glspl{rc}, which can access the elements in the wide registers word by word.
The \glspl{rc} and the \glspl{vwr} are connected through a network of multiplexers, allowing multiple \glspl{rc} to work in parallel on different sections of a \gls{vwr}.

The size of the \glspl{vwr} depends on the target domain, biosignal kernels in our case, and the datapath width consuming the data words.
On the one hand, the \glspl{vwr} need to be large enough to minimize the frequency at which new data have to be loaded.
On the other hand, wider \glspl{vwr} have higher leakage.
In VWRA2, the \glspl{vwr} and the \gls{spm} have a 4096 bitwidth allowing 128~words of 32~bits to be transferred in a single cycle.
This width is a good tradeoff considering that the four \glspl{rc} can access one-fourth of a \gls{vwr} (32 words).
To be single-ported, multiple \glspl{vwr} are needed to store the different operands.
After testing different implementations, we found out that 3 \glspl{vwr} represent a good compromise between performance and energy efficiency.

The \gls{srf} has \num{8} 32-bit entries used for scalar values that are kernel-dependent, such as addresses for the \gls{spm}, masking values for the \glspl{vwr} index computation, or loop parameters for the kernel execution control.
The \gls{srf} is single-ported, allowing one access at a time from the different units (RCs, LSU, MXCU, and LCU).

\subsection{Specialized slots}

We borrow from the VLIW architecture the concept of specialized slots by introducing an LSU, an LCU, and an MXCU on top of the \glspl{rc} of each column, as shown in Fig.~\ref{fig:vwr2a_architecture}.
They all have their own instruction stream (see Table~\ref{tbl:instruction_flow}), synchronized with the \glspl{rc} in the column via the common \gls{pc}.

\subsubsection{LSU: Load-Store Unit}
Controls the data transfers between the SPM and the VWRs or the SRF.
The data arrays are allocated to VWRs while scalar values (e.g., loop size) are stored in the SRF.
The LSU also controls the shuffle unit that is an essential component of VWR2A.
As each RC is accessing only one-fourth of a VWR, data reordering is needed to move the data inside the full VWR.
Such reordering is possible through the RCs connection matrix, but it is highly inefficient in terms of performance and energy, whereas the shuffle unit enables fast and energy-efficient data reordering.
It takes as input the data contained in the VWRs A and B, applies a hardcoded shuffle operation on the data, and stores the result in the VWR C (Fig.~\ref{fig:vwr2a_architecture}).
The available shuffle operations are:

\begin{itemize}
  \item \textit{Words interleaving}: VWR A and B words are interleaved. The result is twice the size of a VWR, and the upper or lower half can be selected as the output.
  \item \textit{Even or odd index pruning}: prunes the even/odd elements of VWRs A and B, and outputs the remaining elements (of both A and B).
  \item \textit{Bit-reversal}: applies bit-reversal shuffling to the concatenation of VWRs A and B. The result is twice the size of a VWR, and the upper or lower half can be selected as the output.
  \item \textit{Circular shift}: the concatenation of VWRs A and B is shifted by 32 words up in a circular manner (i.e., the upper 32 words are moved to the lower 32 words). The result is twice the size of a VWR, and the upper or lower half can be selected as the output.
\end{itemize}

\subsubsection{MXCU: MultipleXer-Control Unit}
Controls the multiplexers that connect the VWRs outputs to the RCs.
Each RC has access to 1/4 of the VWRs width.
To limit the number of control bits, all the RCs access the same address of their slice.
This address is also used to write the data back to any of the VWRs.
Although this structure adds some constraints to the kernel mapping, they can be solved with careful data placement and proper use of the shuffling unit.

\subsubsection{LCU: Loop-Control Unit}
Generates the branches and jumps for the program counter and notifies the synchronizer at the end of a kernel.
It increases the code coverage by allowing the execution of loops with any nest depth and control-intensive code to be efficiently executed on VWR2A.
When multiple columns work in parallel, their respective PCs are synchronized.

\subsection{Kernel mapping}\label{subsec:kernel_mapping}

The FFT kernel mapping is discussed here to illustrate the use of the architecture.
This kernel uses the common in-place \mbox{radix-2} FFT algorithm~\cite{cooley1965}, which reduces the computation complexity from $O(N^2)$ to $O(N\log{N})$, where $N$ is the number of points.
The \mbox{radix-2} algorithm divides the computation into $k$ stages, where $2^k=N$.
All the stages execute the same flow of operations; the only changes are the coefficients and the data ordering.
The shuffle unit applies the ``\textit{words interleaving}'' shuffling to create the correct data layout for the next stage.
Both columns are used to execute the FFT kernel.
The output of the kernel is in bit-reversed order, and the shuffle unit is again used to reorder the data.

An optimized version is used for real-valued FFTs (i.e., the imaginary part is null).
The sequence of $N$ real values is transformed into an $N/2$ complex sequence.
Then, the complex FFT kernel presented above is used.
This technique reduces the computations of the FFT kernel but requires some additional operations, also executed on VWR2A, to recover the correct output.
The overall gain is approximately a factor of~\num{2} compared to a complex FFT of size $N$ where the imaginary part is zero.

These steps are mapped into instructions for the different units (i.e., LCU, LSU, MXCU, and the RCs).
Table~\ref{tbl:instruction_flow} shows a sample of the instruction flow of the different units for the FFT kernel, where $k$ corresponds to the VWRs address that is accessed by the RCs.
The RCs' instructions are grouped for simplicity, but they can all execute a different operation.

\begin{table}[tp]
\caption{VWR2A instruction flow example}
\label{tbl:instruction_flow}
\centering
\begin{tabular}{|@{\hskip0.1cm}c@{\hskip0.1cm}|c|c|c|c|}
    \hline
    PC & LCU & LSU & MXCU & RC0-3 \\
    \hline
    \multicolumn{5}{|c|}{...} \\
    \hline
    3 & NOP & LOAD A & k=0 & NOP \\
    \hline
    4 & i=0 & LOAD B & NOP & NOP \\
    \hline
    5 & i++ & NOP & NOP & VWRC=VWRA+VWRB \\
    \hline
    6 & BLT PC=5 & NOP & k++ & VWRA=VWRA-VWRB \\
    \hline
    \multicolumn{5}{|c|}{...} \\
    \hline
    14 & j++ & STORE A & NOP & NOP \\
    \hline
    15 & NOP & STORE C & NOP & NOP \\
    \hline
    16 & BLT PC=3 & NOP & NOP & NOP \\
    \hline
    17 & EXIT & NOP & NOP & NOP \\
    \hline
\end{tabular}
\end{table}

\section{Experimental setup}\label{sec:expSetup}

\subsection{Biosignal processing ultra-low power embedded platform}

To demonstrate the advantages of our architecture, we have integrated VWR2A in an ultra-low power SoC intended for biomedical signal acquisition and processing~\cite{museicv3}.
This platform features an ARM Cortex M4F processor, \SI{192}{\kibi\byte} of \gls{sram} (divided into six banks that can be individually power gated), an analog front end for the acquisition of biosignals (e.g., \gls{ecg}, \gls{ppg}), a DMA, and several fixed-function accelerators.
The FFT accelerator is used for comparison with our VWR2A because this platform is also oriented to biosignal processing.
It computes FFTs and inverse FFTs up to 4096 points, with an optimized flow for real-valued inputs.
The FFT weights are stored in internal ROMs, whereas a dual-port memory is used to store the data.
To avoid overflow, this custom FFT accelerator uses an internal representation of 18~bits with dynamic scaling.
The SoC elements (e.g., accelerators, memories, processor) are connected through the AMBA-AHB
bus interface.
The SoC has multiple power domains that can be turned on and off during execution to optimize energy consumption further.

\subsection{Integration of our programmable core}

We connected our VWR2A to the AMBA-AHB bus interface, precisely like the other hardware accelerators, to have a fair comparison within the original SoC design.
VWR2A has one master port, controlled by its DMA, to transfer data between the SoC \gls{sram} and the VWR2A \gls{spm}.
The kernel acceleration and DMA transfer requests from the CPU are received through an additional slave port.
VWR2A informs the processor when a kernel execution, or a DMA transfer, is finished through an interrupt line.
VWR2A is included in the same power domain as the other accelerators and can therefore be power gated.

\subsection{Performance and energy characterization}

We synthesized the complete SoC, including VWR2A, with the TSMC \SI{40}{\nano\meter} LP~CMOS technology at \SI{80}{\mega\hertz} (the original SoC frequency) and ran post-synthesis simulations to compare the performance and the energy of both implementations (i.e., custom accelerator versus programmable core).
This allows cycle accurate simulations from which we extract the cells switching activity that we use for power estimation with Synopsys PrimePower~\cite{primepower}.

\subsection{Software benchmarks}

\subsubsection{Standalone kernels}

We first compared our VWR2A with the SoC FFT accelerator using a standalone FFT kernel, which is a typical kernel used in biomedical applications.
We implemented different FFT sizes for both complex and real-valued sequences.
The SoC FFT accelerator uses a mixed \mbox{radix-2} and \mbox{radix-4} implementation.
For our VWR2A, we used the \mbox{radix-2} algorithm presented in Section~\ref{subsec:kernel_mapping}.
The second kernel is a FIR filter with 11 taps.
We used three different input sizes to compare our VWR2A with the CPU.
Our mapping uses two columns of the reconfigurable array that work on different slices of the input array.

\subsubsection{Biosignal application}

To study the impact at the application level of our programmable architecture, we considered the MBioTracker application, which measures cognitive workload~\cite{pale2021MBioTracker}.
This application is divided into four steps: preprocessing, delineation, features extraction, and prediction.
First, the preprocessing applies a FIR filter over the raw input data.
Second, the delineation detects the maximums and minimums of the filtered signal to extract inspiration and expiration times.
Third, these values are used for extraction of time features (mean, median, and RMS values), while the FFT of the filtered signal is employed for frequency features extraction.
Finally, the cognitive workload is estimated using an SVM algorithm.

\section{Experimental results}\label{sec:expResults}

\subsection{Performance on standalone kernels}

\begin{table}[tp]
\caption{FFT kernel performance comparison for various sizes}
\label{tbl:fft_cycle_comparison}
\centering
\begin{tabular}{r@{\hskip0.1cm}|@{\hskip0.1cm}r@{\hskip0.1cm}|@{\hskip0.1cm}r@{\hskip0.15cm}r@{\hskip0.1cm}|@{\hskip0.1cm}r@{\hskip0.15cm}r}
\toprule
& \multicolumn{1}{c|@{\hskip0.1cm}}{\textbf{CPU}} & \multicolumn{2}{c|@{\hskip0.1cm}}{\textbf{FFT ACCEL}} & \multicolumn{2}{c}{\textbf{VWR2A}} \\
\multicolumn{1}{l|@{\hskip0.1cm}}{Complex-valued} & \multicolumn{1}{c|@{\hskip0.1cm}}{cycles} & cycles & speed-up & cycles & speed-up \\
\midrule
\textbf{512} & \num{47926} & \num{7099} & \SI{6.8}{\times} &  \num{7125}  & \SI{6.7}{\times} \\
\textbf{1024} & \num{84753} & \num{13629} & \SI{6.2}{\times} &  \num{12405} & \SI{6.8}{\times} \\
\textbf{2048} & \num{219667} & \num{31299} & \SI{7.0}{\times} &  \num{30217} & \SI{7.3}{\times} \\
\midrule
\multicolumn{1}{l}{Real-valued} & \multicolumn{5}{r}{} \\
\midrule
\textbf{512} & \num{24927} & \num{3523} & \SI{7.1}{\times} &  \num{3666} & \SI{6.8}{\times} \\
\textbf{1024} & \num{62326} & \num{8007} & \SI{7.8}{\times} &  \num{7133} & \SI{8.7}{\times} \\
\textbf{2048} & \num{113489} & \num{16490} & \SI{7.4}{\times} &  \num{14427} & \SI{7.9}{\times} \\
\bottomrule
\end{tabular}
\end{table}

\subsubsection{FFT kernel}

\begin{figure}[tp]
	\centering
	\includegraphics[width=0.52\textwidth]{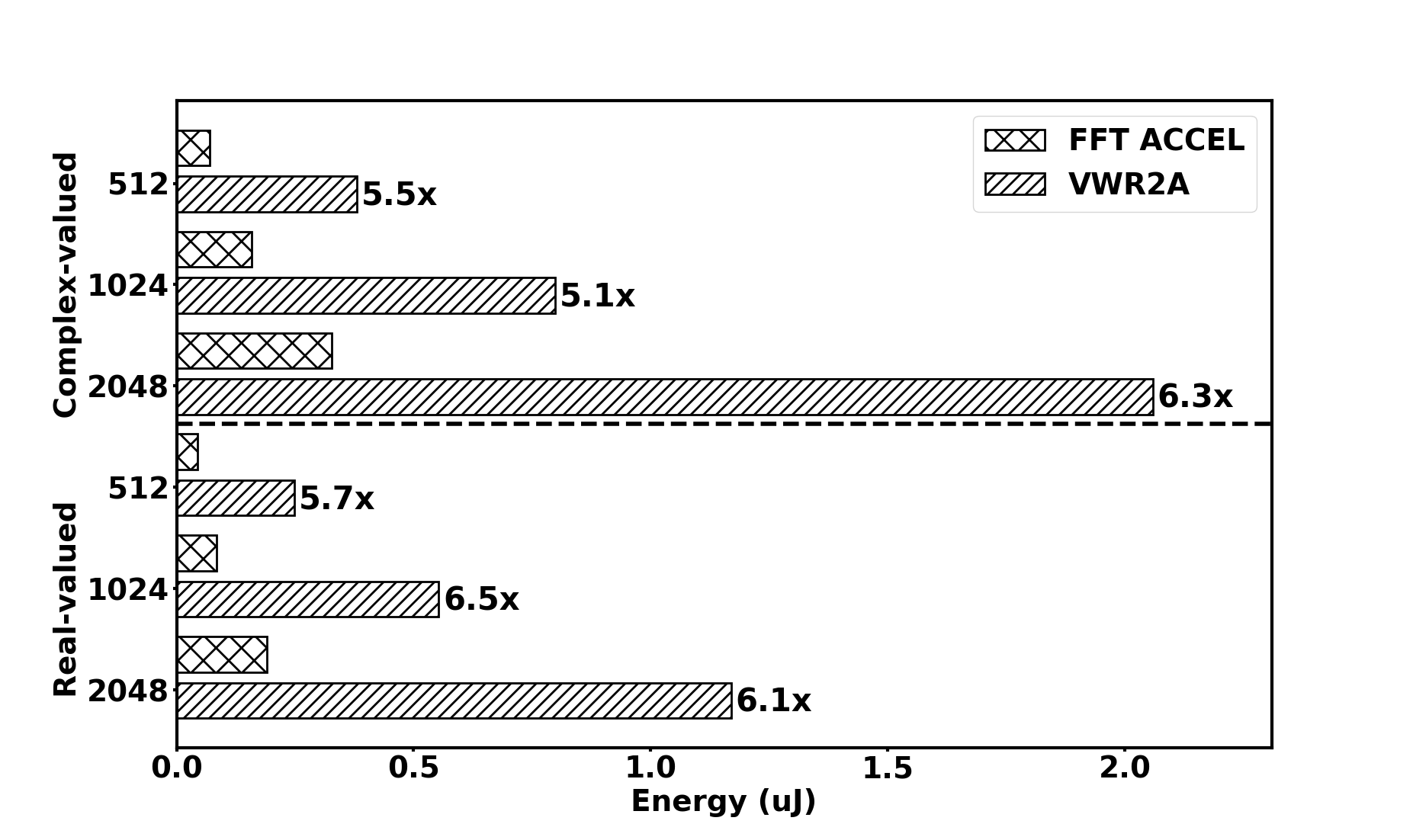}
	\caption{FFT kernel energy comparison for various sizes. Even if the performance of VWR2A is equivalent to that of the custom FFT accelerator, as expected, the gap in energy consumption is still significant in the case of isolated kernels.}
	\label{fig:fft_energy_comparison}
\end{figure}

The performance and energy consumption results for various complex and real-valued FFT sizes are reported in Table~\ref{tbl:fft_cycle_comparison} and Fig.~\ref{fig:fft_energy_comparison}, respectively.
These results show that both the FFT accelerator (FFT ACCEL) and VWR2A have similar performance and are \SI{7.4}{\times} faster than the ARM Cortex-M4 (CPU) on average.
VWR2A is less performant for small FFT sizes because the programming of the DMA transfers and the kernel parameters has a slightly larger overhead than the FFT accelerator programming.
As expected, Fig.~\ref{fig:fft_energy_comparison} shows that the FFT accelerator is more energy-efficient than VWR2A when the specific kernel it was designed for is considered in isolation.
Nevertheless, our goal was to narrow as much as possible the energy gap between both implementations.

The FFT accelerator uses a mixed radix-2 and radix-4 implementation that depends on the actual FFT size, resulting in different performance and power consumption, while the VWR2A mapping is identical for all FFT sizes.
This explains the variation of the energy consumption ratio in Fig.~\ref{fig:fft_energy_comparison}.
Finally, this figure only considers the accelerator energy consumption.
If the complete SoC is considered, the energy difference is between a factor \SIrange{4}{5}{\times}.
In any case, compared to an FFT using only the Cortex-M4 processor and the CMSIS-DSP library with 16-bit data in q15 format, both the FFT accelerator and our architecture produce energy savings, of \SI{86.0}{\percent} and \SI{40.8}{\percent}, respectively.

Table~\ref{tbl:fft_cgra_power} presents the power consumption breakdown per subcomponent, i.e., for the FFT accelerator and VWR2A.
The main contributors in our architecture are the memories and the datapath (i.e., the \glspl{rc}).
This means that the overhead of the instruction control, which is non-negligible in typical instruction-set processors~\cite{b:catthoor2010}, is removed in VWR2A.
The \textit{Memories} category contains the VWR2A \gls{spm} (\SI{32}{\kibi\byte}) and the \glspl{vwr} (\SI{3}{\kibi\byte}), accounting for \SI{46}{\percent} and \SI{54}{\percent} of the total power, respectively.
In contrast, the FFT accelerator has \SI{17}{\kibi\byte} of memory in total.
To build the \gls{spm} wide interface, smaller memory macros of the width supplied by the technology provider were concatenated.
The \glspl{vwr} were built using latches of the standard cell library.
A custom design for these memories will undoubtedly reduce power consumption.
Regarding the \textit{Datapath} category, the available optimizations are more limited because the FFT accelerator is specialized for FFTs, with an 18-bit-wide datapath, while our \gls{cgra} has a more general-purpose 32-bit ALU.
One solution could be to have a 16-bit mode with two simultaneous 16-bit operations instead of one 32-bit operation.

Compared to the Ultra-Low Power Samsung Reconfigurable Processor (ULP-SRP)~\cite{Changmoo2014}, a recent instantiation of the ADRES template that uses the TSMC \SI{40}{\nano\meter} LP technology, VWR2A exhibits significant performance and energy gains.
The authors reported an execution time of \SI{839.1}{\micro\second} and an energy consumption of \SI{19.9}{\micro\joule} for a 256-Point FFT,\footnote{The authors do not specify if they implement a complex-valued or a real-valued FFT. We considered a 256-Point complex-valued FFT, corresponding to the worst case for VWR2A.} while VWR2A executes that same kernel in \SI{35.6}{\micro\second} and consumes \SI{0.3}{\micro\joule}.
These numbers correspond to a factor \SI{23}{\times} improvement in performance and a factor of \SI{66}{\times} in terms of energy.
It is important to note that post-layout simulation has been done for the ULP-SRP, while we ran post-synthesis simulation, which can explain part of the significant difference in energy.

\begin{table}[tp]
\caption{FFT accelerator and VWR2A power breakdown while executing a 512-point real-valued FFT}
\label{tbl:fft_cgra_power}
\centering
\begin{tabular}{l|l@{\hskip0.1cm}r|l@{\hskip0.1cm}r|c}
\toprule
& \multicolumn{2}{c|}{\textbf{FFT ACCEL}} & \multicolumn{2}{c|}{\textbf{VWR2A}} & \\
\textbf{Instance} & Power (\si{\milli\watt}) & \si{\percent} & Power (\si{\milli\watt}) & \si{\percent} & ratio \\
\midrule
\textbf{DMA} & \num{1.07E-02} & \SI{1}{\percent} & \num{9.47E-02} & \SI{2}{\percent} & \num{8.9} \\
\textbf{Memories} & \num{6.68E-01} & \SI{68}{\percent} & \num[retain-zero-exponent]{3.49E-00} & \SI{64}{\percent} & \num{5.2} \\
\textbf{Control} & \num{6.25E-02} & \SI{6}{\percent} & \num{1.00E-01} & \SI{2}{\percent} & \num{1.6} \\
\textbf{Datapath} & \num{2.42E-01} & \SI{25}{\percent} & \num[retain-zero-exponent]{1.72E-00} & \SI{32}{\percent} & \num{7.1} \\
\midrule
\textbf{Total} & \num{9.83E-01} & \SI{100}{\percent} & \num{5.41E-00} & \SI{100}{\percent} & \num{5.5} \\
\bottomrule
\end{tabular}
\end{table}

\subsubsection{FIR filter kernel}

Table~\ref{tbl:fir_filter_table_comparison} reports the experimental results for three different input sizes for the FIR filter with 11~taps.
We compared the performance of the processor (CPU) with that of our VWR2A.
The processor uses the CMSIS-DSP library with 16-bit data (q15 format), while our solution uses 32-bit data.
Table~\ref{tbl:fir_filter_table_comparison} shows that our accelerator is on average \num{14.9} times faster and consumes \SI{71.3}{\percent} less energy than the processor.

\begin{table}[tp]
\caption{FIR filter kernel performance and energy comparison for different numbers of points and 11 taps}
\label{tbl:fir_filter_table_comparison}
\centering
\begin{tabular}{r@{\hskip0.1cm}|@{\hskip0.15cm}r@{\hskip0.1cm}r|@{\hskip0.1cm}r@{\hskip0.15cm}r|@{\hskip0.1cm}r@{\hskip0.15cm}r}
\toprule
 & \multicolumn{2}{c|@{\hskip0.1cm}}{\textbf{CPU}} & \multicolumn{2}{c|@{\hskip0.1cm}}{\textbf{VWR2A}} & \multicolumn{2}{c}{\textbf{GAIN}}\\
& Cycles & Energy & Cycles & Energy & Cycles & Energy \\
& & (\si{\micro\joule}) & & (\si{\micro\joule}) & speed-up & savings \\
\midrule
\textbf{256 pts} & \num{24747} & \num{0.37} & \num{1849} & \num{0.11} & \num{13.4} & \SI{69.9}{\percent} \\
\textbf{512 pts} & \num{49253} & \num{0.73} & \num{3260} & \num{0.21} & \num{15.1} & \SI{71.7}{\percent} \\
\textbf{1024 pts} & \num{98283} & \num{1.45} & \num{6091} & \num{0.40} & \num{16.1} & \SI{72.4}{\percent} \\
\bottomrule
\end{tabular}
\end{table}


\subsection{Performance on biosignal application}

Table~\ref{tbl:biosignal_benchmarks_comparison} reports performance and energy consumption for the different steps of the application.
The complete application has been ported on VWR2A and the processor only manages the high-level control of the application.
This table supports the central claim of this paper and shows that larger savings can be obtained from a reconfigurable architecture with respect to a custom accelerator when complete applications --- rather than individual kernels --- are considered.
The significant gains in performance and energy presented below are due to the parallel computing power of VWR2A (i.e., 8 \glspl{rc} and the specialized slots) and its low power consumption.

\subsubsection{Preprocessing}

The version using our VWR2A is \num{13.2} times faster than the Cortex-M4 (CPU), which translates into energy savings of \SI{64.7}{\percent}.
The CPU+FFT ACCEL version is equivalent to the CPU version because no code can be accelerated by the FFT accelerator (which remains power-gated), hence showing the advantage of a programmable architecture.
This shows the potential benefits that can be obtained by using a programmable architecture that can execute a larger proportion of the overall application.

\subsubsection{Delineation}

This step is a typical example of con\-trol-in\-ten\-sive code.
The computation load is low but there are a lot of \textit{if} conditions used to detect the valid minimums and maximums.
General purpose CPUs are very inefficient at executing such code, while VWR2A can take advantage of its more powerful ILP capabilities.
This translates into a \SI{94.1}{\percent} gain in performance and \SI{82.9}{\percent} savings in energy.
As before, the CPU+FFT ACCEL version is equivalent to the CPU version.

\subsubsection{Features extraction and SVM prediction}

The FFT accelerator computes a real-valued 512-Point FFT, which translates to an \SI{9.3}{\percent} gain in energy compared to the CPU version.
However, the FFT represents only a portion of the application code, of which the custom accelerator cannot execute anything else.
In contrast, VWR2A can execute all the code of the feature extraction step and the SVM prediction, with a corresponding energy saving of \SI{56.0}{\percent} compared to the CPU version.
This result is \SI{6}{\times} better than the CPU+FFT ACCEL version.
VWR2A benefits from its large code coverage, limiting the data transfers between the system's main memory and its SPM.
For example, the FFT uses the filtered data loaded into the SPM during the preprocessing step and keeps the results inside the SPM.
It copies only the estimated state by the SVM back to the system's main memory, while the FFT accelerator has to copy back the 512 FFT output values.

\begin{table}[tp]
\caption{Biosignal application performance and energy comparison}
\label{tbl:biosignal_benchmarks_comparison}
\centering
\begin{tabular}{l|r|r@{\hskip0.1cm}r|r@{\hskip0.1cm}r}
\toprule
& \textbf{CPU} & \multicolumn{2}{c|}{\textbf{CPU +}} & \multicolumn{2}{c}{\textbf{CPU +}} \\
& \textbf{} & \multicolumn{2}{c|}{\textbf{FFT ACCEL}} & \multicolumn{2}{c}{\textbf{VWR2A}} \\
\textbf{Cycles} &  &  & savings &  & savings \\
\midrule
Preprocessing & \num{49760} & \num{49760} & \SI{0.0}{\percent} & \num{3763} & \SI{92.4}{\percent} \\
Delineation & \num{46268} & \num{46268} & \SI{0.0}{\percent} & \num{2723} & \SI{94.1}{\percent} \\
Feat. extraction & \num{70639} & \num{54255} & \SI{23.2}{\percent} & \num{8627} & \SI{87.8}{\percent} \\
\midrule
Total & \num{166667} & \num{150283} & \SI{9.8}{\percent} & \num{15113} & \SI{90.9}{\percent} \\
\midrule
\multicolumn{4}{l}{\textbf{Energy (\si{\micro\joule})}} \\
\midrule
Preprocessing  & \num{0.74}   & \num{0.74} & \SI{0.0}{\percent}  & \num{0.26} & \SI{64.7}{\percent} \\
Delineation & \num{0.74} & \num{0.74} & \SI{0.0}{\percent} & \num{0.13} & \SI{82.9}{\percent} \\
Feat. extraction  & \num{1.1}   & \num{0.98} & \SI{9.3}{\percent}  & \num{0.47} & \SI{56.0}{\percent} \\
\midrule
Total & \num{2.6} & \num{2.5} & \SI{3.9}{\percent} & \num{0.86} & \SI{66.3}{\percent} \\
\bottomrule
\end{tabular}
\end{table}

\section{Conclusion}\label{sec:conclusion}

In this paper, we have shown the feasibility of using a domain-specific programmable core as an accelerator instead of a fixed-function accelerator and achieving similar, or better, performance.
Regarding energy consumption, fixed-function accelerators typically perform better on specific tasks, but when complete applications are considered, our VWR2A has better energy consumption.
The reason is that programmable architectures can accommodate application-specific optimizations that are not possible with custom accelerators.
In addition, more code is also eligible for acceleration with a flexible architecture, which leads to better overall performance and energy efficiency, as long as the energy gap with respect to the fixed-function accelerator has been sufficiently reduced.


\bibliographystyle{ACM-Reference-Format}
\bibliography{references}

\end{document}